# A baseline for multiple-choice testing in the university classroom

**Aaron. D. Slepkov[a]\*, Melissa L. Van Bussel[a,b], Kara. M. Fitze[a,b], and W. S. Burr[b]**

[a]*Department of Physics & Astronomy, Trent University, Peterborough, Canada*
[b]*Department of Mathematics, Trent University, Peterborough, Canada*

\*corresponding author: aaronslepkov@trentu.ca

**Abstract**

There is a broad literature in multiple-choice test development, both in terms of item-writing guidelines, and psychometric functionality as a measurement tool. However, most of the published literature concerns multiple-choice testing in the context of expert-designed high-stakes standardized assessments, with little attention being paid to the use of the technique within non-expert instructor-created classroom examinations. In this work we present a quantitative analysis of a large corpus of multiple-choice tests deployed in the classrooms of a primarily undergraduate university in Canada. Our report aims to establish three related things: First, reporting on the functional and psychometric operation of 182 multiple-choice tests deployed in a variety of courses at all undergraduate levels of education establishes a much-needed baseline for actual as-deployed classroom tests. Second, we motivate and present modified statistical measures—such as item-excluded correlation measures of discrimination and length-normalized measures of reliability—that should serve as useful parameters for future comparisons of classroom test psychometrics. Finally, we use the broad empirical data from our survey of tests to update widely-used item-quality guidelines.

**Introduction**

With dwindling educational resources and increasing student enrollments, multiple-choice testing (MCT) has become a ubiquitous assessment format in higher education. In fact, it is likely that MCT is the most widely used form of assessment in Canadian post-secondary introductory-level courses (DiBattista & Kurzawa, 2011; Slepkov & Godfrey, 2019; Tobias & Raphael, 1996; Suskie, 2018). While the proliferation of MCT has been driven by economic considerations such as the ease and automation of scoring, the technique has remained popular because well-constructed MCT tests prove highly reliable, valid, and fair (Scott *et al.*, 2006; Little & Bjork, 2015; Anderson & Biddle, 1975; Suskie, 2018; Wainer & Thissen, 1993). Much of the (ongoing) developmental research into MCT has been conducted in service of optimizing large high-stakes standardized tests (Haladyna, 2004; Rodriguez, 2005; Moreno *et al.*, 2006), which are now almost exclusively offered in this format. Thus, there has been a large amount of research into ways of improving MCT, including widely-available guidelines on best uses and best practices. Many of these guidelines have been adopted by those studying classroom tests and exams. However, the nature of classroom examination is such that opportunities for maximizing their psychometric attributes (i.e., the functional operation) are limited. This is mostly due to the limited testing times, test lengths, student numbers, and opportunities for iterative test improvements, as compared with high-stakes standardized tests. Additionally, there is a lack of information in the MCT development literature that is targeted towards instructor-designed and test-bank-based classroom tests.

There is a growing awareness of the need to improve classroom MCT (DiBattista & Kurzawa, 2011). Whether researchers are interested in assessing the quality of new experimentally-designed tests or in reviewing the quality of testing in an academic program, such research must be contextualized with respect to typical classroom test attributes. The nature of examinations is such that tests are more shrouded in secrecy and security concerns than other aspects of course instruction. Thus, there is a scarcity of reports on the psychometrics of "typical" classroom MCTs. Many published examples of individual test attributes can be found, but because those studies are invariably reported by assessment experts working to improve testing in a particular academic program, there is likely a strong selection and publication bias in this literature. Thus, a particularly relevant baseline of classroom MCT use would comprise tests made and deployed by average practitioners. A key study that centers the current work was conducted by DiBattista and Kurzawa (2011), where they reported on the functioning of classroom multiple-choice tests in a typical Canadian university. While their work aimed to establish a representative survey of MCT attributes across their institution, they ultimately reported on the psychometric attributes of only 16 tests. To the best of our knowledge, that report remains the broadest publicly-available survey of classroom multiple-choice testing to date.

In this study we present a large survey of test-level and item-level attributes of classroom multiple-choice tests offered at a primarily-undergraduate Canadian university. In contrast to other studies that have looked at individual or small groups of tests, we report on 182 multiple-choice tests that span all undergraduate levels of education and a wide range of academic disciplines. The research objectives of this article are three-fold.

First, we aim to provide a representative sample of traditional multiple-choice tests in the context of higher-education classroom use. Because the analyzed tests are from a wide array of instructors, courses, instructional levels, and disciplines, the primary aim of this work is to provide a representative and useful baseline of classroom MCT psychometrics for future comparisons by practitioners and researchers.

Second, because classroom tests vary widely in operational attributes such as length, the number of test-takers, and quality, the flaws of some commonly-used measures of test quality can become exaggerated and lead to conflicting conclusions. Such statistical drawbacks are often inconsequential for large (and optimized) standardized tests. Alternatively, more sophisticated item analysis tools such as item response theory, Rasch models, and G-theory are often employed in the analysis of large high-stakes standardized tests, but most of these are unlikely to be adopted by classroom test creators. Thus, in this report we aim to shed new light on some drawbacks inherent to commonly-reported classical item analysis methods, and to subsequently offer recommendations for simple modifications and best practices that will facilitate future comparisons between classroom tests. In particular, we discuss the advantages of using item-excluded correlations as measures of item discrimination, and of using length-normalized test reliability parameters.

Third, our two primary objectives align to provide updated and empirically-driven guidelines for assessing the strengths of classroom MCTs. By establishing a representative distribution of item attributes and test psychometrics, we are able to offer data-driven recommendations for what constitutes average, above average, and exceptional measures of item discrimination and test reliability. Furthermore, our survey clarifies longstanding concerns regarding appropriate levels of item difficulty, the prevalence of various option-number items, as well as some presumptive quirks in classroom test-design such as easy "confidence boosters".

**Methods**

*Data Collection*

All the analyzed tests were deployed in midterm or final examinations between years 2013-2019 at Trent University, a small undergraduate-focused institution located in Peterborough, Canada, with approximate undergraduate enrolment of 10,000 and faculty complement of 250. The majority of tests were administered and processed with standard Scantron® forms and optical character recognition software. A few tests were manually entered into a compatible digital format to allow for their inclusion in the study. A wide range of course instructors was solicited to supply raw multiple-choice test data. Potential instructors were initially identified from their use of centralized Scantron tools. Instructors were then contacted by email, requesting authorization of the release of anonymized test data. In an attempt to present the most representative sample of institutional MCT use possible, a handful of tests were solicited from instructors of key disciplines who do not use the university's centralized Scantron processing office. No data was used from instructors who failed to give permission or who subsequently withdrew from the study. No instructor, test-length, quality, or discipline-based criteria were used to exclude data from the survey. Once received, all tests were checked for student anonymity, and anonymized if needed. The majority of collected tests exclusively comprised multiple-choice items. In the rare cases where the multiple-choice section was only a component of the test, the non-multiple-choice items were removed from the analysis and the remainder was treated as a standalone assessment tool (i.e., a test). For this study, we are only concerned with traditional "single-response" MCT that is scored dichotomously, without partial-credit and without penalty for guessing. Some tests included multiple-choice items that fell outside of the standard criteria. These included clear cases of true-false items, multiple-selection items, and items scored with more than one correct option. In this case, the items were likewise excluded from the analysis. In total, 368 items were excluded from the data of 39 tests. Of these, 222 were true-false from 12 tests. In total, we survey the functioning of 182 tests from 45 different instructors, spanning 12 academic disciplines. The data include analysis of a total of 11,246 multiple-choice items and 24,885 student-tests. All item analysis and statistical tests reported herein were conducted in R (version 3.6.3, (R Core Team, 2019)), with custom-written scripts.

### *Item Analysis*

*Item difficulty.* Individual units of evaluation on MCTs are referred to as items. The most common measure of a test item's functionality is that of the item's difficulty. The difficulty index, *p*, is the proportion of students who selected the correct ("keyed") option for a given item. Thus, *p*, which ranges from 0 (no student selected the keyed option) to 1 (all students selected the keyed option), scales inversely with the traditional notion of difficulty and is actually a measure of item facility. The test score, *P*, is the sum of all item difficulties within the test.

*Item Discrimination.* An important aspect of any reliable classroom test is its ability to differentiate between students with strong knowledge of the subject and students with poor knowledge. Ideally, each item gives a small measure of such distinction, and the combination of a group of items allows the test on the whole to discriminate better between more and less knowledgeable students. Under this framework, *item discrimination* is a correlation between item scores and overall ability. The computation of item correlations, especially for longer tests, is quite cumbersome without use of a computer. Thus, having been developed through the pre-computer era, 1930–1970, most historical measures of item discrimination involve approximations and simplifications that vary in computational sophistication. One popular approach, known as the upper-lower discrimination index, *D* (Ebel & Frisbie, 1991; Shete *et al.*, 2015; Allen & Yen, 2001), simply divides the total test scores into an upper percentile and lower percentile grouping (typically upper and lower 27%), and compares the total item scores of these groups, divided by the sample size of the groups. This method remains popular for evaluating classroom test items, largely due to its ease of calculation. A more direct measure of correlation can be found in the *point-biserial correlation*, $r_{pb}$. The point-biserial is the Pearson correlation for dichotomous data, such as traditional multiple-choice items that are scored as zero or one. This is the most widely used measure of test item discrimination, and is typically computed as an "item-total" correlation given by

$$r_{pb} = \frac{\bar{P}_1 - \bar{P}_0}{S_n} \sqrt{\frac{n_1 n_0}{n^2}} \quad (1)$$

and

$$S_n = \sqrt{\frac{1}{n}\sum_i^n (p_i - \bar{P})^2} \qquad (2)$$

where $\bar{P}_1$ and $\bar{P}_0$ are the mean test scores of the group of students who selected the correct and incorrect option on the item, respectively; $S_n$ is the test-score standard deviation; $p_i$ and $\bar{P}$ are the mean scores of item $i$ and mean test score; and $n$, $n_1$, and $n_0$ are the total number of students, and the number of students who selected the correct and incorrect response on the item, respectively.

As an item-total correlation, the point-biserial is somewhat problematic conceptually. This is because the item score makes up part of the total test score and thus adds spurious weight to the total. Mathematically, this means that $\bar{P}$ and $\bar{P}_1$ already contain the information $p_i$ (Allen & Yen, 2001). With widespread access to computers and statistics software, a full calculation of the Pearson correlation between an item and the total test score exclusive of that item can be easily obtained. Such an "*item-excluded correlation*", $r'$, provides an uncontaminated measure of item discrimination. In practice, the overweighed correlation of $r_{pb}$ is never smaller than that of $r'$, with the difference between the two measures decreasing both with increasing test length and discrimination. This fact was known when the point-biserial correlation was developed as a proxy for discrimination (Guilford, 1954; Henrysson, 1963; Cureton, 1966; Zubin, 1934). However, because calculations were conducted by hand, and psychometrics were primarily being developed for large, discriminating, standardized tests, the drawbacks of using item-total correlations were tolerated. For short classroom tests the differences can be sufficiently significant to affect prior published research conclusions. A good illustration of the dangers of reporting $r_{pb}$ instead of $r'$ can be found in considering a hypothetical MCT in which every student blindly guesses on every item: clearly no item on this test is discriminating for any particular knowledge domain. Yet, a calculation of the mean test point-biserial correlation, $\bar{r}_{pb}$, will yield a value of $1/\sqrt{n}$ (Guilford, 1954). By comparison, a calculation of the mean test item-excluded correlation will be zero, as none of the item scores are correlated with any of the other item scores. Given that the typical guideline for evaluating discrimination holds that a value of $r_{pb}$ above 0.20 is considered acceptable (Bodner, 1980; DiBattista & Kurzawa, 2011), a 20-item test comprising entirely randomly-scored items would be considered adequately discriminating when using the point-biserial correlation. Thus, in this work we use the item-excluded score correlation, $r'$, as the primary measure of discrimination. To compute $r'$, we correlate student scores on each item with total test scores that exclude that item. Many statistical packages and Scantron® report formats allow for the calculation of $r_{pb}$ along with $r'$, which is sometimes (for example, in SPSS) referred to as the "corrected item-total correlation" (Varma, 2006).

*Test Reliability.* The consistency with which a test can be used as a tool for accurately ranking student knowledge is known as test-score *reliability* (Frisbie, 1988; Cronbach & Shavelson, 2004). This measure is directly linked to item discrimination; a test comprising more discriminating items is more reliable. Thus, mean item discrimination, $\bar{r}'$ (similarly, $\bar{r}_{pb}$) essentially measures test reliability. Formally, Cronbach's alpha ($\alpha$) is a measure for internal consistency and is the most commonly-reported parameter representing test-retest reliability (Falk & Savalei, 2011; Streiner, 2003). If a test comprises $K$ items, Cronbach's alpha can be computed as

$$\alpha = \frac{K}{K-1}\left(1 - \frac{\sum_{i=1}^{K} \sigma_{p_i}^2}{\sigma_P^2}\right), \tag{3}$$

where $\sigma_P^2$ and $\sigma_{p_i}^2$ are the variances of the test scores and $i^{\text{th}}$ item, respectively. $\alpha$ ranges between 0 (for an entirely random test) to 1 (for a perfectly reliable measure of student ability), and scales with the number of items, such that increasing the number of discriminating items will always improve the test's reliability. Thus, while there are guidelines for what is considered an adequately-reliable test, comparing reliabilities of tests that differ in the number of items can be misleading. A viable means of normalizing Cronbach's alpha to remove the effect of test length involves the transformation of $\alpha$ to that of a scaled test with a standardized number of items. It has become commonplace to use 50 items as the standard for the normalized reliability (Bodner, 1980; DiBattista & Kurzawa, 2011). This is accomplished using the Spearman-Brown prediction formula, which for a test comprising $K$ items normalized to a 50-item test yields

$$\alpha_{50} = \frac{\frac{50}{K}\alpha_K}{1+\left(\frac{50}{K}-1\right)\alpha_K}. \tag{4}$$

In the Recommendations section we use the distribution of obtained test reliabilities to provide updated recommendations for targeted values of $\alpha_{50}$ in classroom examinations.

### *Notation*

As we present the following section with numerical results from our dataset of multiple-choice tests, there are two conventions we will follow. First, when presenting summary statistics, we will present the statistic plus/minus its standard deviation, e.g., 50 ± 12. The standard deviations are intended to give the reader an idea of variability. Secondly, we use the "bra-ket" notation of $\langle \cdot \rangle$ to indicate a mean-of-summaries. That is, many of the quantities computed in this analysis are themselves means of quantities, and we then are often interested in the mean of these means: the bra-ket indicates that the result has been taken as a mean across multiple tests; e.g., $\langle \alpha_{50} \rangle$. For a simple sample mean value we use a bar over the parameter; e.g., $\bar{r}_{pb}$. Table 1 summarizes the key attributes and psychometric measures of the surveyed MCTs, including all quantities that are discussed in the following sections.

## Results and Discussion

### *The multiple-choice testing context*

As university enrollments have exploded in recent years, upper-year courses have grown sufficiently so that the use of MCT is now commonplace across all levels of instruction. Accordingly, the plurality (43%) of tests in our survey come from first-year (i.e., Freshman) courses, over half of the tests come from 2nd- (26%) and 3rd-year (25%) courses (i.e., Sophomore and Junior), and 6% of tests are obtained from fourth-year Senior courses. The abundance of multiple-choice tests at the Sophomore and Junior levels is a reflection of the fact that a larger number of courses are offered at these instructional levels, compared with Freshman courses that are less numerous but are larger and invariably use MCT. Class sizes vary widely across the institution. This fact is reflected in the distribution of the number of students, *n*, responding to the various tests (see Figure 1A). Multiple-choice tests are given in classes as small as 11 students, with 5% of tests given to cohorts smaller than 20 students. In some cases, different versions of the same test are given in consideration of test security. Surprisingly, we find this practice to be common at all class sizes larger than 40 students. Thus, the data shown in Figure 1A includes multiple versions of some tests analyzed as independent tests. The prevalence of multiple-choice testing at the Junior and Senior levels is reflected by the abundance of MCTs taken by cohorts of 40-120 students. Tests of more than 250 students are invariably from introductory-level courses for this corpus.

**Table 1.** Summary measures for 182 classroom multiple-choice tests.

| Measure | parameter | value |
| --- | --- | --- |
| Total # tests surveyed | $n$ | 182 |
| Total # test-takers surveyed | $N$ | 24,885 |
| Total # items surveyed | $K$ | 11,246 |
| Mean ± SD # items | $\bar{K}$ | 62 ± 26 |
| Maximum # items on a test | $K_{max}$ | 106 |
| Minimum # items on a test (or test section) | $K_{min}$ | 17 |
| Maximum # test-takers on a test | $N_{max}$ | 901 |
| Minimum # test-takers on a test (or test version) | $N_{min}$ | 11 |
| Mean ± SD item discrimination | $\langle r' \rangle$ | 0.24 ± 0.17 |
| Mean ± SD test-average item discrimination | $\langle \bar{r}' \rangle$ | 0.24 ± 0.07 |
| Maximum item discrimination | $r'_{max}$ | 0.40 |
| Mean ± SD item difficulty | $\langle p \rangle$ | 0.64 ± 0.22 |
| Mean ± SD test score, as a percentage | $\langle P \rangle$ | 65% ± 8% |
| Mean ± SD adjusted reliability | $\langle \alpha_{50} \rangle$ | 0.77 ± 0.12 |
| Maximum (minimum) adjusted reliability | $\alpha_{50_{max}}(\alpha_{50_{min}})$ | 0.92 (0.20) |
| | | |
| Percentage of poorly discriminating items | $r' < 0$ | 8% |
| Percentage of marginally discriminating items | $0 < r' < 0.15$ | 20% |
| Percentage of adequately discriminating items | $0.15 < r' < 0.35$ | 45% |
| Percentage of well-discriminating items | $r' > 0.35$ | 25% |
| Percentage of items with undefined discrimination | $p = 0,1$ | 2% |

Classroom multiple-choice tests also vary in terms of length, as can be seen in Figure 1B. In practice, the number of items often reflects the length of the test in time, with final examinations that span 2-3 hours employing more items than midterm examinations that span 1-2 hours. However, the correlation between test length in time and number of items is not strong: while surveyed tests with more than 50 items were almost exclusively final examinations, many final examinations comprised fewer than 40 questions. As seen in Figure 1B, test sizes ranged from less than 20 items to more than 100 items. Our data collection protocols are blind to cases where the multiple-choice component is only one part of the total examination. Nonetheless, all surveyed tests with fewer than 50 items were deployed with Scantron® cards that were auto-scored by computer. The mean number of multiple-choice items in a given test is 62 ± 22 (mean ± standard deviation), with the smallest MCT component being 17 items, and the largest being 106.

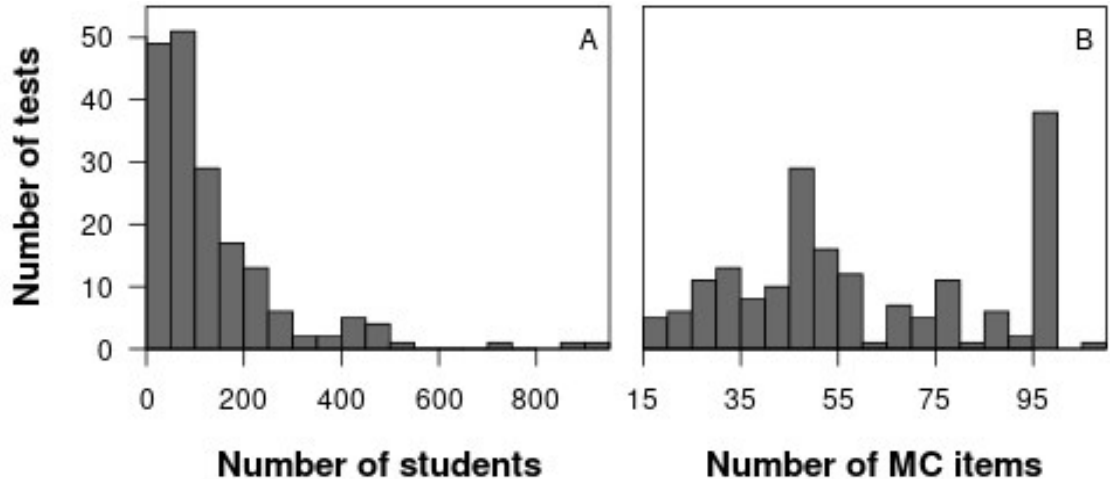

*Figure 1.* Distribution of test cohort size and test length. (A) distribution of number of students per test. (B) distribution of the number of multiple-choice items per test.

### *Prevalence of 3, 4, and 5–option items*

Establishing the optimal number of options offered within MTC items has been an active area of research (Rodriguez, 2005; Raymond *et al.*, 2019; Owen & Froman, 1987). Traditionally, the use of between 3 and 5 options has been most common. Deciding on the number of options to offer in a classroom examination represents a balance of considerations, as the primary consideration for offering more options is a desire to mitigate the effects of student guessing. In practice, this desire is beset by the difficulty of writing large numbers of viable distractors (non-keyed options). The disadvantage of deploying nonfunctional distractors (broadly defined by Raymond *et al.* as options that elicit negligible attention and prove non-discriminating (2019)) is that they take up both time and cognitive space in the test and risk lowering both test validity and score reliability. Studies of the optimal number of options consistently find that the 3-option format is best; at least from a psychometric standpoint (Rodriguez, 2005). However, despite best-practice for the exclusive use of 3-option items, 4- and 5-option items are commonly offered by educational publisher test banks, and the distribution of option-number item types in instructor-created classroom tests remains undiscussed in the literature.

Our content-agnostic data collection method precludes an absolute determination of item scoring rules for every item. For example, a single student who selects a non-offered option (such as selecting F in a 5-option A–E item) is sufficient to mis-identify the item as a 6-option item when examined anonymously, as we did. Nonetheless, we are able to positively code the vast majority of test items in the survey. Among all items, 31% are found to be of the 5-option type, 56% are 4-option, and only 6% are of 3-option, with 7% of items remaining uncategorized or miscoded. Furthermore, most instructors appear to opt for heterogeneous mixed-type tests. We define a test as homogeneous (in terms of option-number type) if the mode type represents over 90% of the total. Based on this definition, we find that 40% (71 of 182) of the tests are homogeneous, and 60% (111 of 182) are heterogeneous. Of the homogeneous tests, 65% comprise 4-option items and 35% comprise 5-option items; that is, none use 3-option items. In fact, not only is the 3-option format unpopular in homogeneous tests, it also does not comprise a significant proportion of mixed-type tests. Within heterogeneous tests, 32% of items are 5-option, 50% are 4-option, and only 8% are 3-option (with the remainder 10% being of ambiguous categorization). Thus, there is a clear and persistent gap between recommendations for best-practices in high-stakes MCTs and in-practice deployment of classroom tests.

### *Item difficulty and test scores*

All classroom tests are a blend of questions of varying difficulty. When considering the opportunity for guessing inherent in MCT, the psychometrically-optimal item difficulty should be near the midpoint between the expectation for guessing and a perfect score (Lord, 1953; Allen and Yen, 2001; Doran, 1980). Thus, depending on the number of options available for an item (e.g., 3, 4, or 5), the optimal item difficulty should be in the range of 0.60-0.67 (midpoints ranging from 0.20 to 0.33; perfect score being 1.0). Across all items, our surveyed mean item difficulty is $\bar{p} = 0.64 \pm 0.22$, which is indeed in the optimal range. However, as shown in Figure 2, the distribution of item difficulty in classroom tests is such that it spans the entire range from 0 to 1 and is skewed towards more easy items of $p > 0.6$. We also find an overabundance of extremely easy questions. In fact, items for which over 90% of students chose the correct response comprise nearly 15% of all test items. This contrasts with the case of validated standardized tests that avoid use of items that are overly easy, and therefore are of limited discrimination.

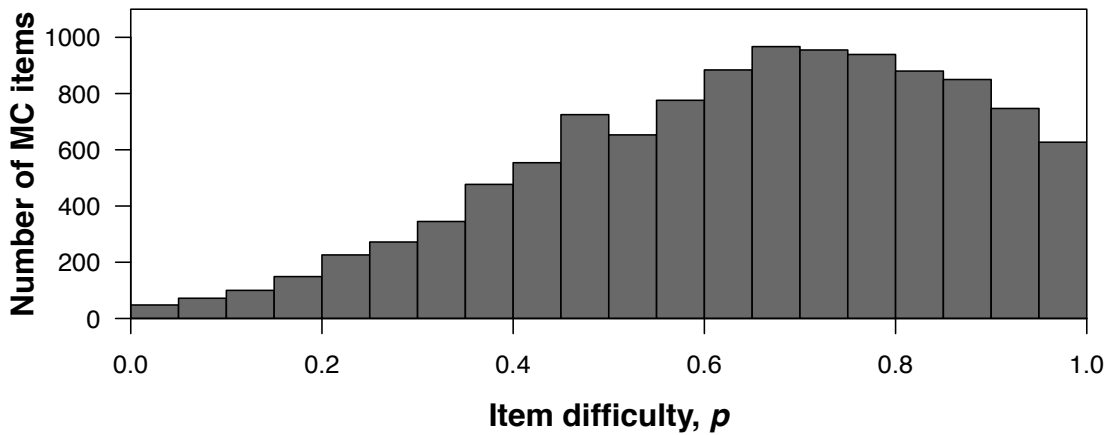

*Figure 2. Distribution of test item difficulty, p, across a total of 11,246 items. p = 0 is maximal item difficulty (no student selected the correct response) and p = 1 is minimal difficulty (all students selected the correct response).*

Most university courses—particularly at the introductory level—are surveys of topics rather than sequences of culminating knowledge. Thus, it may not be expected that item difficulty should increase (*p* decrease) sequentially throughout the test. In discussing test-design with various course instructors we have often noticed that the first few questions are sometimes particularly easy. There is a general sense that this arises by design, often as an attempt to "boost confidence" in test takers. This practice is reflected in our data. Figure 3 presents the mean item difficulty as a function of position within the test, averaged across all tests. In order to test the notion that the first position has an idiosyncratic facility, the first two items are averaged separately from the rest of the test, which is then divided by length into tenths. The analysis confirms that the first few multiple-choice items are generally much easier than the average test item. This lends credence to the notion that—whether knowingly or not—many instructors are placing confidence boosters at the beginning of tests. Furthermore, the trend in Figure 3 suggests that tests tend to get more difficult as they progress, but the effect size is very small because the mean item difficulty varies considerably more between the various tests than it does as a function of position within a test.

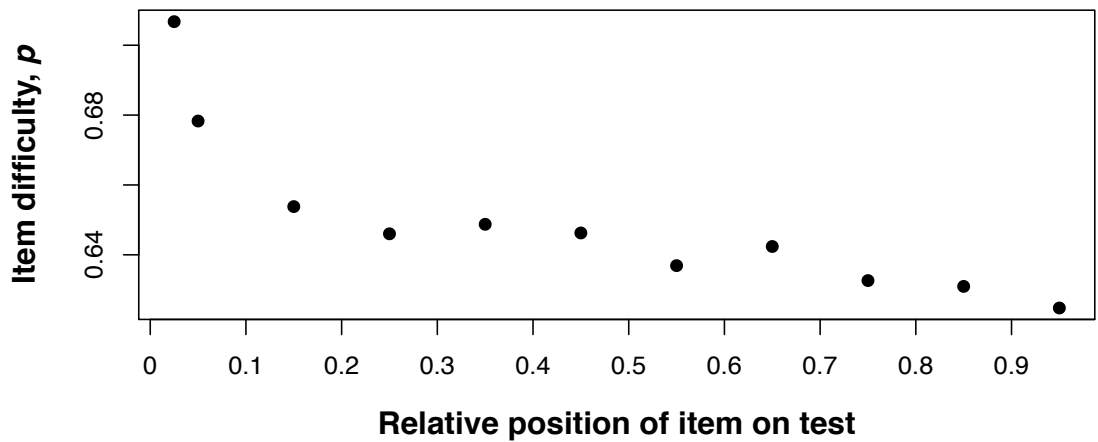

**Figure 3**. *Progression of mean item difficulty with in MCT. The relative position axis indicates the location of the question relative to the test length, with all values after 0.1 representing binned average item difficulty across positions covering 0.1 of the relative position. Recall that p decreases with increasing item difficulty.*

Anticipated test scores are a primary consideration in classroom test design. Because item difficulties can span the full range of possibilities, it is likely that many test makers adjust the composition of their tests in an effort to attain a target test score. However, the success of such design hinges on instructors' ability to know or predict the difficulty level of test items. As seen in the distribution of test scores shown in Figure 4, most of the surveyed tests show a class average in the range of 60%-70%. The mean test score is 65% ± 8%, and the most common test score is nominally 65%. The range of test averages is large, spanning 42% to 89%. A "pass" in the Canadian post-secondary system is 50%.

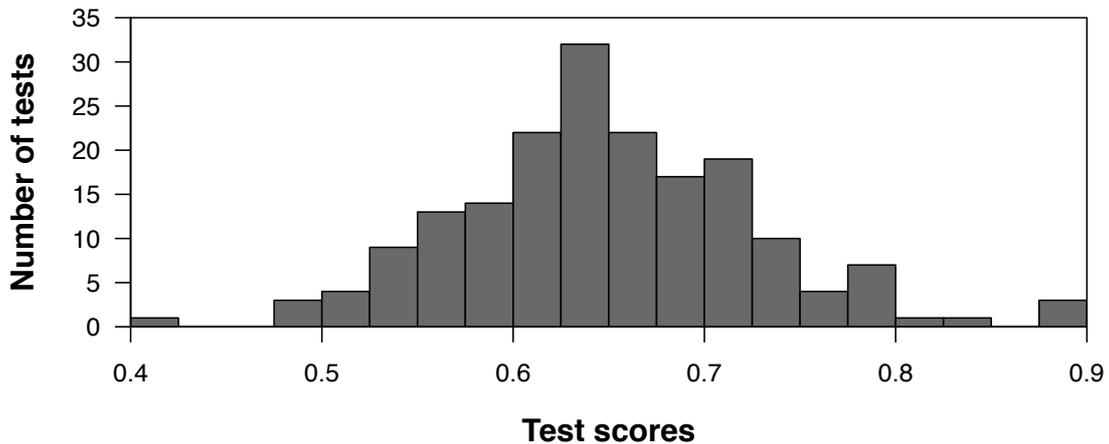

*Figure 4.* Distribution of average test-scores across a total of 182 classroom multiple-choice tests.

### Item Discrimination

Figure 5 displays the distribution of individual item discrimination for our survey, both in terms of the conventional $r_{pb}$ and $r'$. As expected, the distribution of the (item-total) $r_{pb}$ values is similar to that of (the item-excluded) $r'$, with the latter appearing to be simply shifted downwards. The mean value of $r_{pb}$ across the 11,246 items is $0.29 \pm 0.17$. The mean value of $r'$ is $0.24 \pm 0.17$. Note that the difference between the two means of the distributions does not imply that the conversion of $r_{pb}$ to $r'$ is constant. Rather, the difference between the two measures depends both on the number of items in a test and strength of the correlation (discrimination) itself. Overall, $r_{pb} - r' = 0.059 \pm 0.025$. From the distribution of item discrimination, we find that a relatively wide range of values is observed in practice, ranging as high as $r' = 0.62$ and as low as $-0.62$. Recommendations for what can be considered acceptable or excellent in terms of item discrimination are somewhat fluid (see the Recommendations section). However, as a measure of discrimination, the interpretation of $r' < 0$ is unambiguous: negatively-discriminating items undermine the test as a reliable measurement tool. Unfortunately, as others have pointed out previously (DiBattista & Kurzawa, 2011; Tarrant *et al.*, 2006; Downing, 2005; Sayyah *et al.*, 2012), highly flawed, negatively discriminating items are relatively common in classroom examinations. Indeed, we find that 8% of the surveyed items had $r' < 0$. There are several reasons why an item might display negative discrimination, wherein more knowledgeable students select the keyed response at a lower rate than do less-knowledgeable students. One possibility is that a particular distractor becomes increasingly plausible as one applies a greater scope of knowledge but that such knowledge is not being tested in the item. Thus, top students may be "overthinking" the question. A distractor analysis in which the discriminating power of each distractor is measured would be useful in such cases. Just as the keyed response should strongly correlate with the overall test scores, the distractors should anti-correlate (i.e., give $r' < 0$). Thus, any distractors that significantly discriminate positively should be reconsidered by the test-maker. Nonetheless, in our experience, the improper keying of an item is the most common culprit in negatively-discriminating classroom test items.

The mean of item discrimination for each test provides a useful descriptive measure of test quality. The distribution of mean item discrimination for the 182 tests, both in terms of $\bar{r}_{pb}$ and $\bar{r}'$ are shown in Figure 6. We find a wide range of mean test discrimination, with some tests yielding $\bar{r}'$ as low as 0.05 and as high as 0.40, with a mean value across all tests of 0.24 ± 0.07. In Figure 7 we examine the mean test item discrimination, $\bar{r}'$, by year of course code (Freshman through Senior), and see that while the distribution of mean item discrimination does decrease somewhat from first to third year, those tests offered to seniors were *better* tests than the majority of the first-year tests. This contrasts with the prevailing notion of MCT being a poor substitute for in-depth written classroom testing instruments, and indicates that upper-year tests may prove overall equally discriminating to those given in lower-year courses. This finding is somewhat counter-intuitive. One of the longstanding criticisms of multiple-choice testing is that it prioritizes the testing of low-level factual and categorical knowledge over more complex application, analysis, and synthesis knowledge. One would presume that introductory-level courses assess knowledge at lower levels of complexity than do Senior courses. Thus, it might be expected that, within the context of classroom tests, multiple-choice testing is more suitable for lower-level courses. Our results indicate that regardless of the suitability (i.e., validity) of MCTs as a function of instruction level, the tests prove only marginally less discriminating in practice at the upper levels, and possibly more discriminating under certain circumstances. In particular, selection bias may be responsible for this finding, wherein those instructors most comfortable with (or expert in) MCT deploy the technique at upper levels.

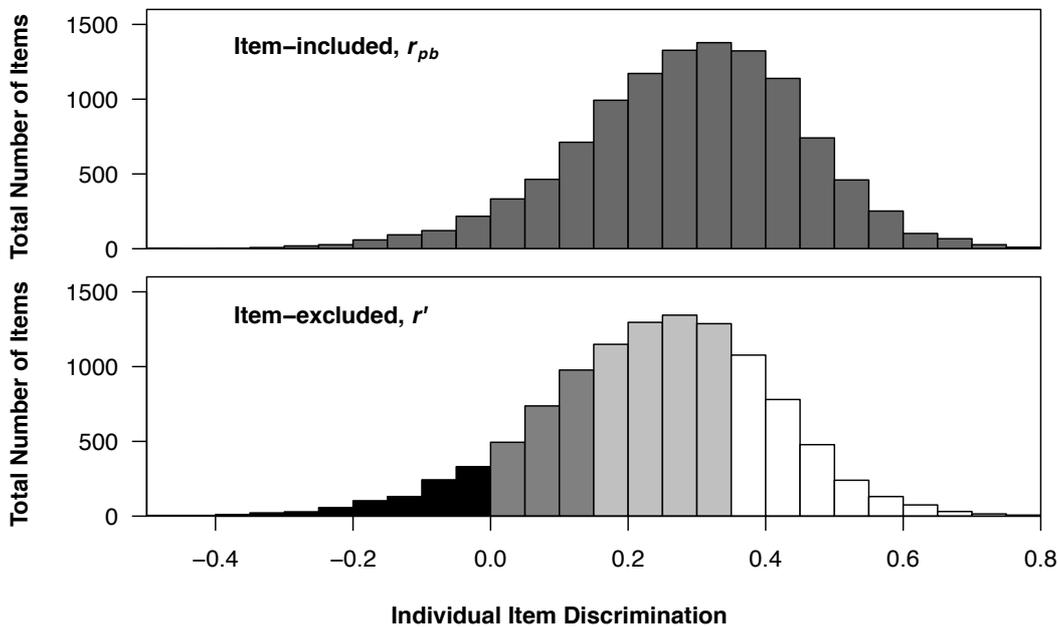

*Figure 5*. Distribution of individual item discrimination for 11,246 classroom test items. Top: point-biserial correlation as a traditional measure of item discrimination. Bottom: Item-excluded correlation as a calibrated measure of item discrimination. Bottom panel includes shading based on four recommended categories for item quality: Poor, Marginal, Adequate, and Excellent.

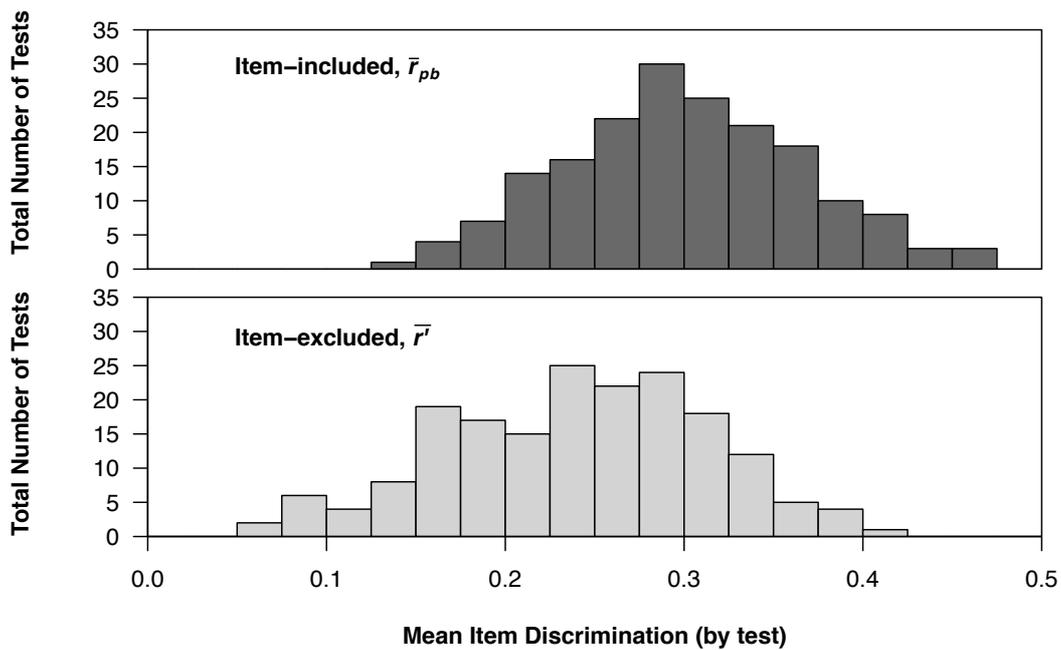

*Figure 6.* Distribution of mean item discrimination for 182 classroom tests. Top: point-biserial correlation as a traditional measure of item discrimination. Bottom: Item-excluded correlation as a measure of item discrimination.

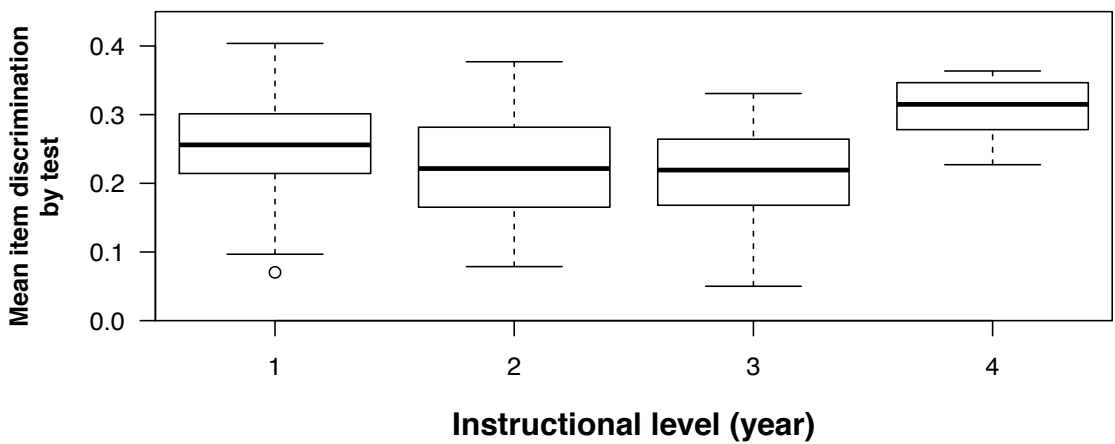

*Figure 7.* Mean test item discrimination, $\bar{r}'$, for classroom tests deployed across the 4-year instruction levels. Data presented as boxplots, with the dark center line representing the median mean test item discrimination. The single dot below the boxplot for year 1 is an outlier, and all other data points are encompassed by the range bars.

In practice, item difficulty and item discrimination are not wholly independent. Theoretically, in the absence of guessing, items with a difficulty of $p = 0.5$ should prove overall most discriminating (Allen & Yen, 2001; Sim & Rasaih, 2006). On the other hand, items approaching the extremes of difficulty/facility have restricted opportunity for discrimination because all student responses are the same in those cases. Our results confirm such variation in item discrimination as a function of item difficulty. Figure 8 presents the mean item-excluded discrimination averaged across all items by ranges of item difficulty. Consistent with the findings of DiBattista & Kurzawa (2011), items in the difficulty range of $0.4 - 0.9$ prove most discriminating on average. The difference in mean item discrimination in this range is very small, making such items almost equally viable. A steep decline is observed for $p < 0.3$, reaching a mean discrimination of nearly zero for very difficult questions with $p < 0.1$. Interestingly, the decline is not as steep for very easy questions, and some items with $p > 0.9$ remain somewhat discriminating. This asymmetry is not surprising, but rather reflects that item difficulties are not expected to naturally fall below the guessing threshold of $0.2 \leq p \leq 0.33$, depending on the number of options in the items (5 through 3). Others have noted that "easy" items (with $p \approx 0.9$) routinely yield acceptably discriminating behaviour (Su *et al.*, 2009; Ebel & Frisbie, 1991). As noted above, we found that a disproportionate number of easy ($p > 0.8$) questions were used in classroom tests (Figure 2). Given that, on average, easy questions do not overly suffer from poor discrimination, this practice does not prove particularly detrimental.

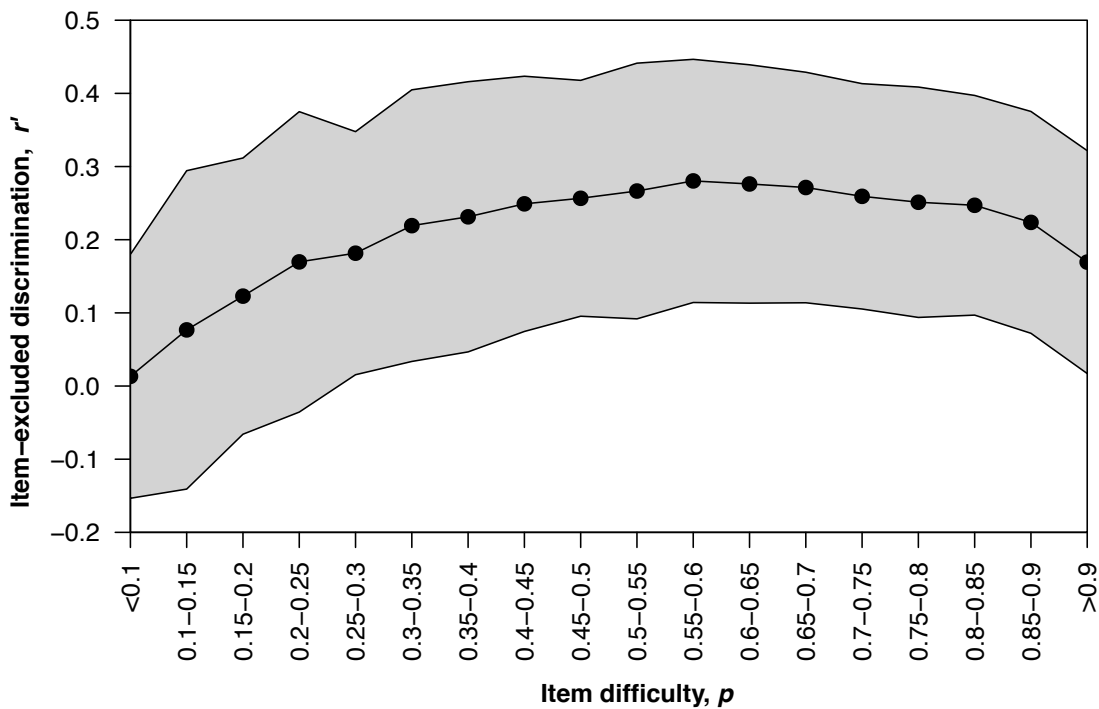

*Figure 8*. Item discrimination and item difficulty. Dots are the mean item discrimination at each item difficulty level, and grey shading represents the standard deviation (not a confidence interval).

### *Test-score reliability*

Test-score reliability is observed to vary widely among the surveyed tests. The mean value of reliability for our tests is $\langle \alpha \rangle = 0.78 \pm 0.12$, and spans values from 0.29 to 0.94. Because reliability scales (asymptotically) with the number of items, a large portion of this variability comes from the wide range of test lengths. For some of the surveyed tests, the Scantron-based multiple choice section only represents a portion of the total examination, and thus the reliability of such a portion to fully test the desired knowledge is not required to be particularly high. With *tests* ranging from 17 to 106 items, there is an expectation that a wide range of values of Cronbach's $\alpha$ would be found. However, because reliability is an important measure of test quality and thus of the quality of the underlying test items, it is most reasonable to compare length-adjusted reliability. When adjusted to a normalized length of 50 items, the mean reliability value observed is $\langle \alpha_{50} \rangle = 0.77 \pm 0.12$, and spans values from 0.20 to 0.92. This result is an excellent match to previously reported results from Frisbie (1988), where the author found that 50 classroom chemistry tests had $\langle \alpha_{50} \rangle = 0.78 \pm 0.08$.

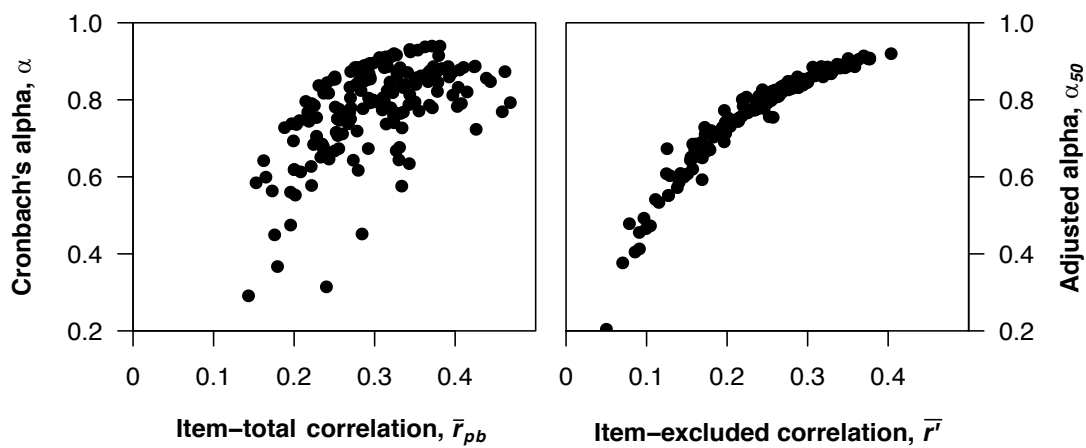

*Figure 9*. Test-score reliability and mean item discrimination. Left: Cronbach's $\alpha$ is only weakly correlated with the uncorrected point biserial coefficient, $\bar{r}_{pb}$. Right: Length-adjusted reliability, $\alpha_{50}$ is highly correlated with the mean item-excluded discrimination coefficient, $\bar{r}'$. This figure (right) allows for a coarse conversion between $\alpha_{50}$ values and item-excluded correlations in previously published works.

It is currently fashionable (see, e.g., Frisbie, 1988; Ebel & Frisbie, 1991; Tavakol & Dennick, 2011) to gauge the quality of a test by considering various ad-hoc guidelines for test reliability. Several recommendations, for instance, suggest that a value of $\alpha$ in the range of 0.7–0.85 constitutes an acceptable-to-good test, while a value greater than 0.90 implies an excellent test. However, $\alpha$ is not meant to represent test quality, but rather is a specific measure of test-score reliability (Frisbie, 1988). It is an absolute and standalone measure that is not particularly useful for comparisons between tests of unequal lengths. Certainly, to be valid, any summative test must have a semblance of reliability. But, precisely how reliable a (say) 25-item 1-hour midterm exam needs to be before it is jettisoned as "unreliable" is not psychometrically defined. Most often, when used in the context of classroom tests, reliability is used to compare whole-test quality among experimental variants. To this end, the normalized reliability, $\alpha_{50}$ is far more appropriate. However, test reliability is closely related to item-discrimination (Ebel 1967). The relationship between the two parameters is typically obscured by how both parameters can scale with the number of items. For example, as shown in Figure 9, the correlation between the non-normalized reliability, $\alpha$, and the test-mean point-biserial correlation, $\bar{r}_{pb}$, is weak. On the other hand, the correlation between the length-adjusted reliability, $\alpha_{50}$, and the item-excluded mean correlation, $\bar{r}'$, is extremely strong. Furthermore, as seen in the figure, the mean item-excluded correlation grows faster than does the reliability, which is bounded above by 1.0 but readily reaches values above 0.8 for many classroom tests.

### *Distractor Analysis and Study Drawbacks*

Much of the recent effort in analyzing the quality of instructor (or test bank) generated MCTs is dedicated to analysis of distractor functioning (DiBattista & Kurzawa, 2011; Tarrant *et al*., 2009; Hingorjo & Jaleel, 2012; Testa *et al*., 2018; Ware & Vik, 2009). The functionality of a set of distractors is simultaneously tied to a distractor's ability to garner significant student selection, and for its ability to adequately discriminate for student ability. That is, a good distractor elicits significant student response and displays adequate negative discrimination. The primary interest in analysis of discriminator functionality in classroom tests lies in a dual desire to improve test efficiency by minimizing superfluous reading and by improving item discrimination. Items with fully functioning distractors are likely to be more discriminating than items with one or more non-functioning distractors (DiBattista & Kurzawa, 2011; Tarrant *et al*., 2009; Hingorjo & Jaleel, 2012). The lack of discriminator analysis in this work is a minor drawback of the current study. Because our data collection is ignorant of the actual test content, situations arise where we cannot distinguish between non-functioning terminal distractors and the heterogeneous distribution of the number of options for a given item. For example, in a test that is mostly made up of 4-option (A-D) items, there are occasions where no test-takers respond to the item with the last option (i.e., option D). In that case we cannot distinguish between a non-functioning distractor (option D) and the possibility that the item has only three options (A-C). Certainly, **as reported by others, we find that the last option is disproportionately non-functioning**, but without absolute knowledge of the number of options in every test item, we cannot provide a quantitative distractor analysis. However, because distractor analysis is ultimately of interest in the context of improving item discrimination and test reliability (Attali & Fraenkel, 2000), the psychometric analysis provided in this report suffices for the evaluation of item and test quality.

Our approach to data collection was guided by a desire to establish a representative survey of the multiple-choice testing in our institution. However, once data collection was completed, we became aware of a large academic department that processes their own multiple-choice tests, rather than using the centralized university Scantron processing office. We then solicited instructors from that department, eventually obtaining data for an additional 14 of the tests included in our analysis. Furthermore, the 182 tests represent data from only 45 distinct instructors and 61 distinct course codes. It is highly likely that many students completed multiple tests in our survey, but we cannot quantify this because of how the data has been anonymized. Overall, our survey provides a good, if imperfect, representation of the use and functionality of multiple-choice testing in our institution and should prove highly useful for contextualizing future MCT research.

**Recommendations**

Over the years, numerous recommendations and guidelines have been offered for interpreting multiple-choice item and test psychometrics. Most such recommendations are provided with an eye to maximize the quality of MCT items and are thus beneficial and informative for classroom test designers. However, because most of the recommendations are either given in the context of professional standardized tests or are based on theoretical psychometric considerations, they are not particularly suitable for classroom test design. On the other hand, our broad survey of classroom MCTs provides a unique context and opportunity to construct useful guidelines from representative data for the Canadian post-secondary education system, and may also be of interest to other similarly structured education systems elsewhere. Thus, in what follows, we provide recommendations and guidelines for the interpretation of item difficulty, item discrimination, and overall test quality/reliability.

Most recommendations for item difficulty emphasize that the greatest opportunity for discrimination exists for items with a value of $p$ midway between that of 1 and the expectation value for guessing. In practice we find that a wide range of item difficulty values provide nearly equal opportunity for item discrimination. Specifically, items with values of $p$ that are within $\approx 0.10$ of either 1 or of the expectation value for guessing provide excellent opportunities for adequate item discrimination. Thus, for 4-option items, for example, targeted item difficulties should be between $p = 0.35 - 0.90$. Items with scores outside of this range are likely to suffer from reduced discrimination.

A common guideline for interpreting item discrimination breaks the range of $r_{pb}$ into value categories (Bodner, 1980; Doran, 1980): Values below zero are poor; the items are significantly flawed or miss-keyed and should be replaced in future iterations. Items with $r_{pb}$ between 0 and 0.20 are considered marginally discriminating and are recommended to be revised or jettisoned. Items displaying $r_{pb} > 0.20$ are considered adequately discriminating, and those above 0.40 are considered excellent. According to such guidelines, we find that of the $\approx 11{,}000$ items in our survey, 5% are poor, 22% are marginal, 46% are adequate, and 25% are excellent in discrimination. These findings are in broad agreement with those of 16 classroom tests analyzed by DiBattista and Kurzawa (2011), albeit more items in our survey are found to show good discrimination. However, as mentioned above, because the item-total point-biserial correlation becomes inflated for items within few-item tests, measuring discrimination via the item-excluded correlation index, $r'$ is far more suitable for comparing item discrimination between tests of different lengths. Thus, we provide modified category ranges for the interpretation of this statistic. The mean difference between $r_{pb}$ and $r'$ is found to be 0.059, and this value thus sets a natural choice for how to modify the ranges of $r'$. However, choosing ranges that are instead reduced by 0.05 is more convenient and less esoteric. Negative values for discrimination provide the same conceptual interpretation for both $r_{pb}$ and for $r'$, and thus the definition of a poorly discriminating item should remain unchanged. Specifically, in our survey we find that 8% of the items are poor ($r' < 0$); 20% are marginal ($0 < r' < 0.15$); 45% are adequate ($0.15 < r' < 0.35$); and 25% are well-discriminating, or excellent ($r' > 0.35$). This guideline is summarized at the bottom of Table 1. Note that while the interpretation of the meaning of a negatively-discrimination item is the same, because $r'$ is invariably smaller than $r_{pb}$ more items are identified as truly poor when conducting a proper, item-excluded, discrimination analysis.

Guidelines for the interpretation of Cronbach's $\alpha$ are somewhat arbitrary. Ultimately, $\alpha$ is a measure of the robustness of the test scores to repeated measurement, but different amounts of score uncertainty can be tolerated depending on the purpose of the test. This is to say that an acceptably reliable classroom midterm exam may prove to be woefully unreliable as a high-stakes standardized test. Although standardized tests aim for $\alpha > 0.90$, such tests comprise manifold more items than a classroom test. Classroom test scores are known to span a wide range of reliability, likely averaging below 0.6 (Frisbie, 1988). The standard recommendation for classroom tests is to attain a reliability of at least 0.7 (Downing, 2004). In practice, the reporting of reliability in the classroom test development literature is done more as a proxy for overall test quality, rather than strictly as a measure of internal test-score reliability. In that case, it often makes more sense to compare a whole-test measure of quality that controls for the number of items. To this end, the adjusted reliability, $\alpha_{50}$, is a useful tool. As a point of reference, we find that of the 182 tests, 22% have $\alpha_{50} < 0.7$; 27% have $0.7 < \alpha_{50} < 0.8$; 47% have $0.8 < \alpha_{50} < 0.9$; and 4% have $\alpha_{50} > 0.9$. These ranges are adequate for assessing test quality as poor, marginal, good, and excellent, respectively. That is, classroom tests with $\alpha_{50} > 0.8$ are appropriate for summative or ranking purposes. As shown in Figure 9, the mean item-excluded discrimination value, $\bar{r}'$ is strongly correlated with the adjusted reliability, and thus could replace it as a whole-test measure of quality. Because we recommend using $r'$ as the best measure of individual item quality, we likewise recommend $\bar{r}'$ as a convenient and useful measure of test quality. The asymptotic nature of reliability means that significant improvements in quality of already excellent tests will be reflected by small increases in $\alpha_{50}$. On the other hand, $\bar{r}'$ increases rapidly with test quality and is thus a more useful and intuitive statistic. From the data in Figure 9 we can establish guidelines for the interpretation of $\bar{r}'$ based on the aforementioned benchmarks in $\alpha_{50}$: $\bar{r}' < 0.15$ is poor-to-marginal; $0.15 < \bar{r}' < 0.25$ is adequate; $0.25 < \bar{r}' < 0.35$ is good; and $\bar{r}' > 0.35$ is excellent. Incidentally, we find that of the 182 surveyed tests from Trent University 20 are poor-to-marginal, 76 are adequate, 76 are good, and 10 are excellent.

**Summary and Outlook**

We have presented a broad examination of multiple-choice testing "on the ground" at an undergraduate-education focused Canadian university. Our dataset is the largest of its kind and has allowed for comparisons between typical theoretically-driven recommendations for item analysis and empirical findings for as-deployed classroom tests and examinations. This study thus presents an opportunity for establishing a baseline for a wealth of future research into the development, strengthening, and assessment of MCT in the tertiary education setting.

Several empirical findings are of particular interest:

First, expert recommendations for the preferential use of 3-option items are entirely unheeded by classroom test designers. We find that in practice, 4-option items are most popular, followed by 5-option items. 3-option items comprise less than 10% of all deployed MCT items. It is likely that the preference for 4- and 5-option items stems from a combination of the desire to mitigate successful guessing and a dearth of 3-option items in publisher-made test bank questions.

Second, a number of attributes that are theoretically considered to be deficiencies in test design appear to be of limited impact in practice. Particularly, across 11,246 items and 182 tests, item difficulty was largely concentrated above 0.6. despite this, test averages and overall student performance did not suffer. Thus, it does not seem to be of high importance for instructors to emphasize the difficulty of their questions, so long as *p* falls roughly in the range from 0.4 to 0.9. Rather than being counter-indicated, the use of "easy" ($p > 0.8$) questions is actually supported by item discrimination values that remain sufficiently large.

Third, the presumptive practice of using confidence boosting "easy" ($p > 0.9$) questions at the start of tests is observed. This practice does not appear to degrade the psychometric functionality of the tests. We did not attempt to measure the psychosocial impact of these confidence boosters on overall performance, but the mean difficulty difference of approximately 0.05 units was not as large as might be assumed from common understanding.

Fourth, use of item-included item-total correlations, such as the point-biserial, to measure discrimination metrics is demonstrated to inflate individual item discrimination scores, as was understood by theoreticians, but perhaps not well understood by practitioners. This inflation is large enough to distort the mean test-level examination of test quality through use of average item discrimination, and we strongly recommend against further use of this statistic, especially when using tests with fewer than 100 items. Instead, an item-excluded correlation provides a cleaner measure of discrimination that can be used for comparing items between tests of different lengths. Many modern item-analysis computer programs provide the option for calculating item-excluded correlations.

Fifth, this large set of solicited MCTs from a typical Canadian university proves to have surprisingly good performance, with $\langle \bar{r}' \rangle = 0.24 \pm 0.07$. Furthermore, 47% of the surveyed tests yield mean item discrimination measures deemed either good or excellent. This is encouraging given the lack of formal training or introspection that most instructors are able to devote to design and analysis of their testing instruments. In practice, most instructors are happy with course-level test averages and variances in acceptable ranges, and rarely examine individual item-level discrimination.

Finally, when examining the reliability of MCTs in the tertiary education setting, we strongly advocate for the use of normalized $\alpha$ – such as $\alpha_{50}$ – as it is both valid as a cross-test comparator, and also is highly correlated with the item-excluded mean correlation, $\bar{r}'$. Ultimately, because $\bar{r}'$ is so strongly correlated to $\alpha_{50}$ and furthermore displays a better dynamic range of values, we recommend it be used as the better whole-test statistic.

In conclusion, we leave the reader with a list of quantitative recommendations for the use of evaluating test reliability and efficacy. Classroom test designers and practitioners should aim for the following:

(1) Individual item difficulty ranging from 0.35 to 0.90, with the highest discrimination occurring for difficulties in the $0.55 - 0.70$ range.

(2) Minimum individual item discrimination of 0.15, aiming for 0.35 ($0.15 - 0.35$ being 'good', more than 0.35 being excellent) when using item-excluded correlation, $r'$.

(3) Minimum test-level mean item discrimination, $\bar{r}'$, of 0.25, aiming for 0.35 (0.25 − 0.35 as 'good', more than 0.35 as excellent).

(4) Score reliability with normalized Cronbach's alpha of $\alpha_{50} > 0.8$.